\documentstyle[12pt,epsfig]{article}

\begin{document}
\begin{titlepage}
\begin{center}

{\Large\bf Longitudinal Polarized Parton Densities Updated}

\end{center}
\vskip 2cm
\begin{center}
{\bf Elliot Leader}\\
{\it Imperial College London\\ Prince Consort Road, London SW7
2BW, England }
\vskip 0.5cm
{\bf Aleksander V. Sidorov}\\
{\it Bogoliubov Theoretical Laboratory\\
Joint Institute for Nuclear Research, 141980 Dubna, Russia }
\vskip 0.5cm
{\bf Dimiter B. Stamenov \\
{\it Institute for Nuclear Research and Nuclear Energy\\
Bulgarian Academy of Sciences\\
Blvd. Tsarigradsko Chaussee 72, Sofia 1784, Bulgaria }}
\end{center}

\vskip 0.3cm
\begin{abstract}
\hskip -5mm

We have re-analyzed the world data on inclusive polarized DIS, in
both NLO and LO QCD, including the new HERMES and COMPASS data.
The updated NLO polarized densities are given in both the $\rm
\overline{MS}$ and JET schemes. The impact of the new data on the
results is discussed.


\end{abstract}

\end{titlepage}

\newpage
\setcounter{page}{1}

Since the famous European Muon Collaboration (EMC) experiment
\cite{EMC} at CERN in 1987, substantial efforts, both
experimental and theoretical, have been devoted to understanding
the partonic spin structure of the nucleon, {\it i.e.}, how the
nucleon spin is built up out from the intrinsic spin and orbital
angular momentum of its constituents, quarks and gluons. Our
present knowledge about the spin structure of the nucleon comes
mainly from polarized inclusive and semi-inclusive DIS experiments
at SLAC, CERN, DESY and JLab, polarized proton-proton collisions
at RHIC and polarized photoproduction experiments. The
determination of the longitudinal polarized parton densities in
QCD is one of the important and best studied aspects of this
knowledge.

In this Brief Report we present an updated version of our Set 1
and Set 2 NLO QCD polarized parton densities in both the $\rm
\overline{MS}$ and the JET (or so-called chirally invariant)
\cite{JET} factorization schemes, as well as the LO ones,
determined in our recent analysis \cite{JHEP}. Comparing to our
previous analysis: i) The old HERMES proton \cite{HERMESp} and
neutron \cite{HERMESn} data are replaced with the new HERMES
proton and deuteron data \cite{newHERMES} and ii) a complete
treatment of the recent COMPASS data on the longitudinal
asymmetry $A_1^d$ \cite{COMPASS} is included. A FORTRAN package of
the obtained polarized parton densities (LSS'05) will be
presented at the Durham HEPDATA web site to be used for practical
purposes.\\

In QCD the spin structure function $g_1$ has the following form
($Q^2 >> \Lambda^2$):
\begin{equation}
g_1(x, Q^2) = g_1(x, Q^2)_{\rm LT} + g_1(x, Q^2)_{\rm HT}~,
\label{g1QCD}
\end{equation}
where "LT" denotes the leading twist ($\tau=2$) contribution to
$g_1$, while "HT" denotes the contribution to $g_1$ arising from
QCD operators of higher twist, namely $\tau \geq 3$. In Eq.
(\ref{g1QCD}) (the nucleon target label N is dropped)
\begin{equation}
g_1(x, Q^2)_{\rm LT}= g_1(x, Q^2)_{\rm pQCD} + h^{\rm TMC}(x,
Q^2)/Q^2 + {\cal O}(M^4/Q^4)~, \label{g1LT}
\end{equation}
where $g_1(x, Q^2)_{\rm pQCD}$ is the well known (logarithmic in
$Q^2$) pQCD contribution and $h^{\rm TMC}(x, Q^2)$ are the
calculable {\cite{TMC} kinematic target mass corrections, which
effectively belong to the LT term. In Eq. (\ref{g1QCD})
\begin{equation}
g_1(x, Q^2)_{\rm HT}= h(x, Q^2)/Q^2 + {\cal O}(\Lambda^4/Q^4)~,
\label{HTQCD}
\end{equation}
where $h(x, Q^2)$ are the {\it dynamical} higher twist ($\tau=3$
and $\tau=4$) corrections to $g_1$, which are related to
multi-parton correlations in the nucleon. The latter are
non-perturbative effects and cannot be calculated without using
models.

Let us recall that the Set 1 polarized parton densities
correspond to fits to $g_1/F_1$ and $A_1(\approx g_1/F_1)$ data
(so called '$g_1/F_1$' fits):
\begin{equation}
\left[{g_1(x,Q^2)\over
F_1(x,Q^2)}\right]_{exp}~\Leftrightarrow~{g_1(x,Q^2)_{\rm
pQCD}\over F_1(x,Q^2)_{\rm pQCD}}~, \label{A1HT}
\end{equation}
where for the structure functions $g_1$ and $F_1$ their leading
twist NLO QCD expressions are used. (Why this method is incorrect
for extracting the LO polarized PDs, is discussed in Ref.
\cite{LSS_PRD}.) The Set 2 polarized PD correspond to fits to
$g_1/F_1$ and $A_1$ data where the experimental data for the
unpolarized structure function $F_1(x,Q^2)$ are used
\begin{equation}
\left[{g_1^N(x,Q^2)\over F_1^N(x,
Q^2)}\right]_{exp}~\Leftrightarrow~ {{g_1^N(x,Q^2)_{\rm
LT}+h^N(x)/Q^2}\over F_1^N(x,Q^2)_{exp}}~. \label{g1F2Rht}
\end{equation}
\vskip 0.4cm

As usual, $F_1$ is replaced by its expression in terms of the
usually extracted from unpolarized DIS experiments $F_2$ and $R$
and phenomenological parametrizations of the experimental data
for $F_2(x,Q^2)$ \cite{NMC} and the ratio $R(x,Q^2)$ of the
longitudinal to transverse $\gamma N$ cross-sections \cite{R1998}
are used. Note that such a procedure is equivalent to a fit to
$g_1$ data themselves and we will refer to these as '$(g_1^{\rm
LT}+ \rm HT)$' fits. In this case the HT corrections to $g_1$
cannot be compensated and have to be taken into account (for more
details see our previous paper \cite{JHEP}). In (\ref{g1F2Rht})
$g_1^N(x,Q^2)_{\rm LT}$ (N=p, n, d) is given by the leading twist
expression (\ref{g1LT}) in LO/NLO approximation including the
target mass corrections and $h^N(x)$ are the dynamical $\tau=3$
and $\tau=4$ HT corrections which  are extracted in a {\it model
independent way}.

As in our previous analyses \cite{JHEP,LSS2001}, for the input LO
and NLO polarized parton densities at $Q^2_0=1~GeV^2$ we have
adopted a simple parametrization
\begin{eqnarray}
\nonumber
x\Delta u_v(x,Q^2_0)&=&\eta_u A_ux^{a_u}xu_v(x,Q^2_0),\\
\nonumber
x\Delta d_v(x,Q^2_0)&=&\eta_d A_dx^{a_d}xd_v(x,Q^2_0),\\
\nonumber
x\Delta s(x,Q^2_0)&=&\eta_s A_sx^{a_s}xs(x,Q^2_0),\\
x\Delta G(x,Q^2_0)&=&\eta_g A_gx^{a_g}xG(x,Q^2_0),
\label{inputPPD}
\end{eqnarray}
where on the RHS of (\ref{inputPPD}) we have used the MRST98
(central gluon) \cite{MRST98} and MRST99 (central gluon)
\cite{MRST99} parametrizations for the LO and NLO($\rm
\overline{MS}$) unpolarized densities, respectively. The
normalization factors $A_i$ in (\ref{inputPPD}) are fixed such
that $\eta_{i}$ are the first moments of the polarized densities.
To fit better the data in LO QCD, an additional factor $(1+
\gamma_v x)$ on the RHS is used for the valence quarks. Bearing
in mind that the light quark sea densities $\Delta\bar{u}$ and
$\Delta\bar{d}$ cannot, in principle, be determined from the
present inclusive data (in the absence of polararized charged
current neutrino experiments) we have adopted the convention of a
flavor symmetric sea
\begin{equation}
\Delta u_{sea}=\Delta\bar{u}=\Delta d_{sea}=\Delta\bar{d}= \Delta
s=\Delta\bar{s}. \label{SU3sea}
\end{equation}
Note that this convention affects the results for the valence
parton densities, but not the results for the strange sea quark a
gluon densities.

The first moments of the valence quark densities $\eta_u$ and
$\eta_d$ are constrained by the sum rules
\begin{equation}
a_3=g_{A}=\rm {F+D}=1.2670~\pm~0.0035~~\cite{PDG}, \label{ga}
\end{equation}
\begin{equation}
a_8=3\rm {F-D}=0.585~\pm~0.025~~\cite{AAC00}, \label{3FD}
\end{equation}
where $a_3$ and $a_8$ are non-singlet combinations of the first
moments of the polarized parton densities corresponding to
$3^{\rm rd}$ and $8^{\rm th}$ components of the axial vector
Cabibbo current
\begin{equation}
a_3 = (\Delta u+\Delta\bar{u})(Q^2) - (\Delta
d+\Delta\bar{d})(Q^2)~, \label{a3ga}
\end{equation}
\begin{equation}
a_8 =  (\Delta u +\Delta\bar{u})(Q^2) + (\Delta d +
\Delta\bar{d})(Q^2) - 2(\Delta s+\Delta\bar{s})(Q^2)~. \label{a8}
\end{equation}

As in \cite{JHEP}, we have used the MRST02(NLO) unpolarized parton
densities \cite{MRST02} to constrain via positivity our Set 1 and
Set 2 polarized PD in both $\rm \overline{MS}$ and JET schemes.\\

The numerical results of our fits to the world data \cite{EMC,
newHERMES,COMPASS,world} on $g_1/F_1$ and $A_1$ are presented in
Tables I and II. The data used (190 experimental points) cover the
following kinematic region $\{0.005 \leq x \leq 0.75,~~1< Q^2
\leq 58~GeV^2\}$. The total (statistical and systematic) errors
are taken into account. The systematic errors are added
quadratically. It is seen from the Tables I and II that the values
of $~\chi^2/{\rm DF}(\rm \overline{MS})~$ and $~\chi^2/{\rm
DF}(\rm JET)$ coincide almost exactly for the '$g_1/F_1$' as well
as for the $(g_1^{\rm LT}+\rm HT)$ fits, which is a good
indication of the stability of the analysis regardless of the
scheme used. Analytic formulae for the input parton densities and
higher twist contributions are given in the Appendix.

Let us note the main features of our new results:

~~{\it i}) The new Set 1 and Set 2 polarized PD are close to
those determined in our recent analysis \cite{JHEP}. However, due
to more accurate HERMES/d and COMPASS (small $x$ region) data the
polarized PD are better determined now (see the errors in the
Tables 1 and 2 in \cite{JHEP} and this paper).

~{\it ii}) Compared to our previous results \cite{LSS_PRD,LSS2001}
we obtain now smaller values for the gluon polarization (the first
moment of $\Delta G(x, Q^2))$, which leads to a smaller difference
between the values of the strange quark polarization (the first
moment of $\Delta s(x, Q^2)$) determined in the $\rm \overline
{MS}$ and the JET schemes, respectively. As a consequence, the
difference between the quark helicity $\Delta \Sigma$ determined
in these two schemes is also smaller. This fact has been already
observed in our recent analysis \cite{JHEP}.

{\it iii}) The shape of both the polarized strange quark and gluon
densities is rather different from that of LSS'01 \cite{LSS2001}
(see Fig. 1). This is mainly due to the different positivity
bounds imposed on the polarized PD in the LSS'01 and LSS'05 fits.
The impact of positivity constrains on polarized PD has been
discussed in detail in \cite{JHEP}.
\begin{figure}[bht]
\centerline{ \epsfxsize=2.2in\epsfbox{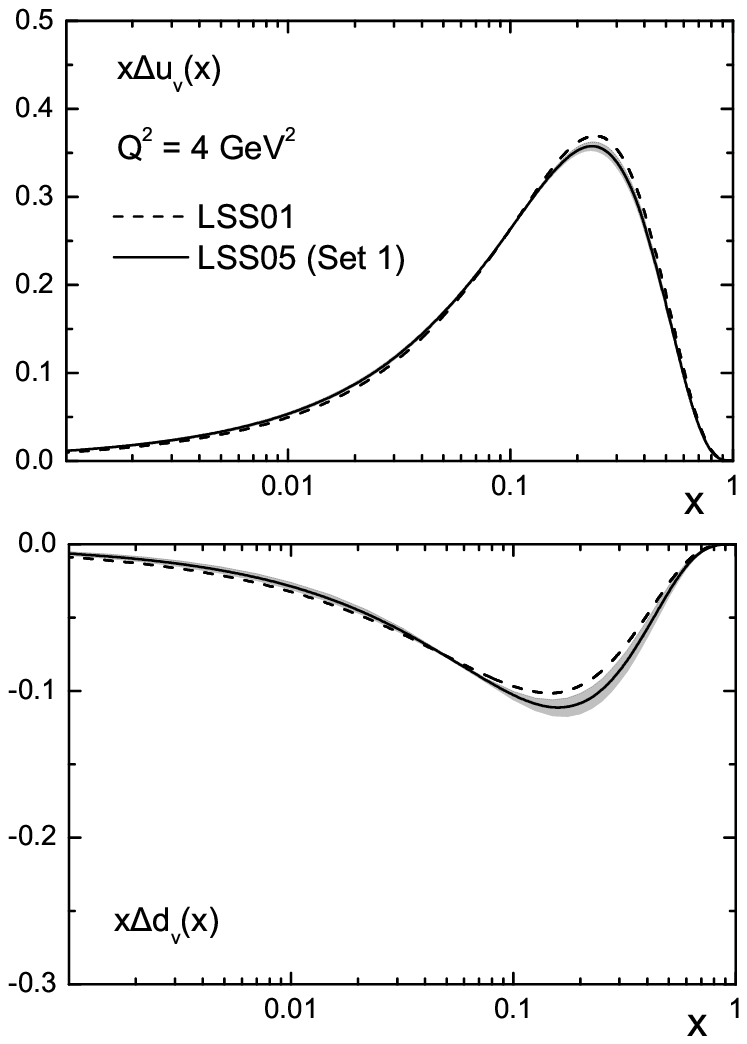}
\epsfxsize=2.2in\epsfbox{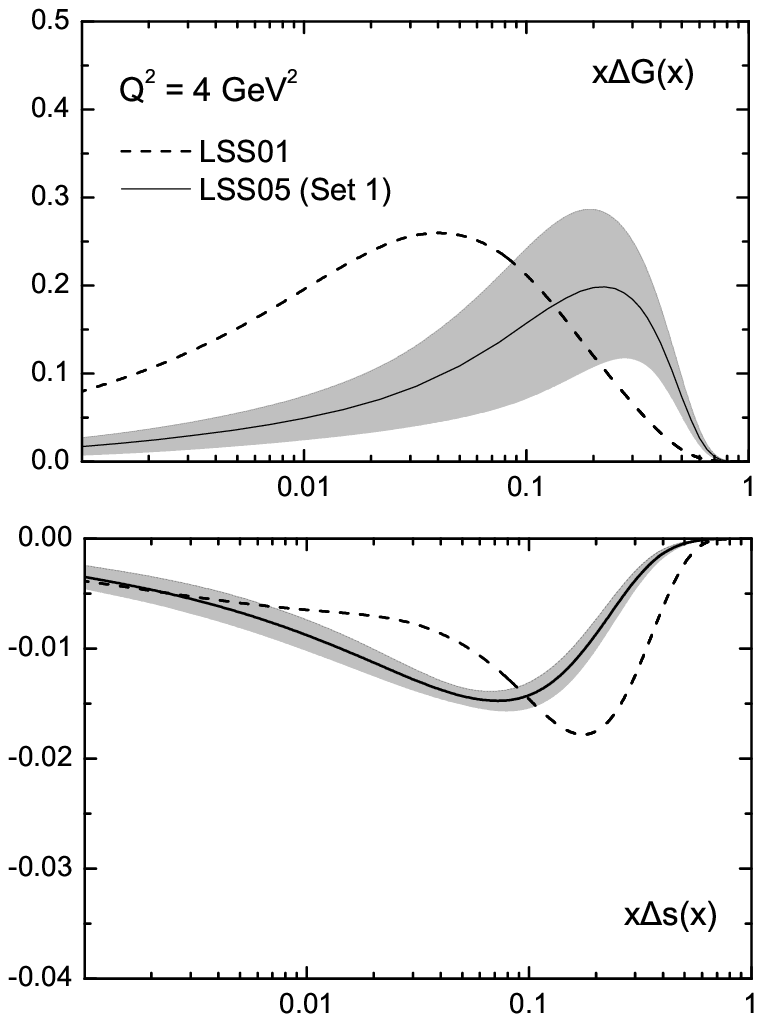} }
 \caption{
Comparison between our two sets of NLO(${\rm \overline {MS}}$)
polarized parton densities, LSS'01 and LSS'05(Set 1), at
$Q^2=4~GeV^2$.
\label{inter1}}
\end{figure}

~{\it iv}) The effect of the new data on the higher twist
corrections to the proton and neutron spin structure functions,
$h^p(x)$ and $h^n(x)$, is negligible (see Fig. 2). The new values
are in good agreement with the old ones although there is a
tendency for the central values for the proton target to be
slightly higher than the old ones. For the neutron target the
only difference is that the new value of $h^n(x)$ at $x \approx
0.2$ is considerably higher than the old one and definitely
different from zero.

Finally, let us illustrate how important the higher twist
corrections are for the description of the data. In Fig. 3 we
compare the new very accurate HERMES deuteron data on the ratio
$g_1/F_1$ at measured $x$ and $Q^2$ with the theoretical curves
for $g_1$ obtained by: i) the best fit to the data using {\it
only} the LT term in Eq. \ref{g1F2Rht} (dotted curve) and ii) the
best fit to the data taking into account in (\ref{g1F2Rht}) the
HT corrections too (solid curve). The dashed curve corresponds to
the LT term when the HT corrections are accounted for. As
expected, the role of HT corrections is significant in the small
$x$ region where the values of $Q^2$ are relatively small: $Q^2
\approx 1.2-2.5~GeV^2$.
\begin{figure}[bht]
\centerline{ \epsfxsize=2.2in\epsfbox{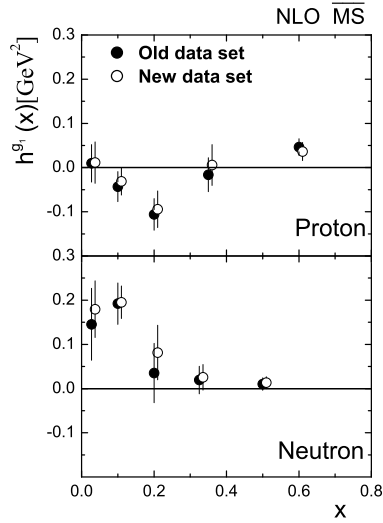}
 }
 \caption{
Effect of the new data on the higher twist values.
 \label{inter2}}
\end{figure}
\begin{figure}[bht]
\centerline{ \epsfxsize=3.2in\epsfbox{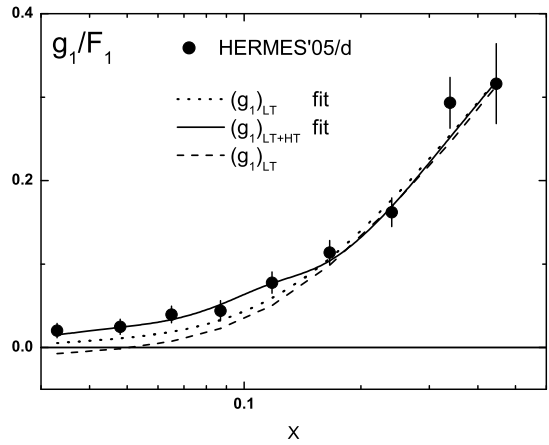} } \caption{ The
effect of higher twist corrections (see the text). \label{inter3}}
\end{figure}
\vskip 6mm { \bf Acknowledgments} \vskip 4mm This research was
supported by the UK Royal Society and the JINR-Bulgaria
Collaborative Grants, and by the RFBR (No 05-01-00992,
03-02-16816).

\newpage
\vskip 10mm {\Large \bf Appendix} \vskip 6mm

{\bf A. Input Parton densities}

For practical purposes we present here explicitly our Set 1 and
Set 2 of polarized parton densities at $Q^2=1~\rm GeV^2$. The
polarized valence quark densities correspond to the convention of
a SU(3) flavour symmetric sea.  \\

LSS'05 ({\bf Set 1}) - NLO($\rm \overline {MS}$) PD($g_1/F_1$):
\setcounter{equation}{0}
\renewcommand\theequation{A.\arabic{equation}}
\begin{eqnarray}
\nonumber x\Delta
u_v(x)&=&~~0.4604~x^{0.6240}~(1-x)^{3.428}~(~1+2.179~x^{1/2}+
14.57~x~)~,\\
\nonumber x\Delta
d_v(x)&=&-0.02408~x^{0.3680}~(1-x)^{3.864}~(~1+35.47~x^{1/2}+
28.97~x~)~,\\
\nonumber x\Delta
s(x)&=&-0.02226~x^{0.3546}~(1-x)^{7.649}~(~1+3.656~x^{1/2}+
19.50~x~)~,\\
x\Delta G(x)&=&~~641.0~x^{3.341}~(1-x)^{6.879}~(~1-3.147~x^{1/2}+
3.148~x~)~. \label{Set1NLOMS}
\end{eqnarray}
\vskip 0.6cm

LSS'05 ({\bf Set 1}) - NLO(JET) PD($g_1/F_1$):
\begin{eqnarray}
\nonumber x\Delta
u_v(x)&=&~~0.4589~x^{0.6224}~(1-x)^{3.428}~(~1+2.179~x^{1/2}+
14.57~x~)~,\\
\nonumber x\Delta
d_v(x)&=&-0.02490~x^{0.3813}~(1-x)^{3.864}~(~1+35.47~x^{1/2}+
28.97~x~)~,\\
\nonumber x\Delta
s(x)&=&-0.01577~x^{0.3127}~(1-x)^{7.649}~(~1+3.656~x^{1/2}+
19.50~x~)~,\\
x\Delta G(x)&=&~~923.6~x^{4.138}~(1-x)^{6.879}~(~1-3.147~x^{1/2}+
3.148~x~)~.
\label{Set1NLOJET}
\end{eqnarray}
\vskip 0.6cm

LSS'05 ({\bf Set 2}) - LO PD($g_1 + \rm HT$):
\begin{eqnarray}
\nonumber x\Delta
u_v(x)&=&~~0.1760~x^{0.3012}~(1-x)^{3.177}~(1+1.610~x)(~1-0.4085~x^{1/2}+
17.60~x~)~,\\
\nonumber x\Delta
d_v(x)&=&-0.00765~x^{0.1535}~(1-x)^{3.398}~(1+3.797~x)(~1+37.25~x^{1/2}+
31.14~x~)~,\\
\nonumber x\Delta
s(x)&=&-0.04660~x^{0.3542}~(1-x)^{8.653}~(~1-0.9052~x^{1/2}+
11.53~x~)~,\\
x\Delta G(x)&=&~~303.6~x^{3.188}~(1-x)^{6.879}~(~1-3.147~x^{1/2}+
3.148~x~)~.
\label{Set2LO}
\end{eqnarray}
\vskip 0.6cm

LSS'05 ({\bf Set 2}) - NLO($\rm \overline {MS}$) PD($g_1 + \rm
HT$):
\begin{eqnarray}
\nonumber x\Delta
u_v(x)&=&~~0.5041~x^{0.6689}~(1-x)^{3.428}~(~1+2.179~x^{1/2}+
14.57~x~)~,\\
\nonumber x\Delta
d_v(x)&=&-0.02847~x^{0.4364}~(1-x)^{3.864}~(~1+35.47~x^{1/2}+
28.97~x~)~,\\
\nonumber x\Delta
s(x)&=&-0.02759~x^{0.3740}~(1-x)^{7.649}~(~1+3.656~x^{1/2}+
19.50~x~)~,\\
x\Delta G(x)&=&~~303.6~x^{3.188}~(1-x)^{6.879}~(~1-3.147~x^{1/2}+
3.148~x~)~. \label{Set2NLOMS}
\end{eqnarray}
\vskip 0.6cm

LSS'05 ({\bf Set 2}) - NLO(JET) PD($g_1 + \rm HT$):
\begin{eqnarray}
\nonumber x\Delta
u_v(x)&=&~~0.4991~x^{0.6639}~(1-x)^{3.428}~(~1+2.179~x^{1/2}+
14.57~x~)~,\\
\nonumber x\Delta
d_v(x)&=&-0.02693~x^{0.4133}~(1-x)^{3.864}~(~1+35.47~x^{1/2}+
28.97~x~)~,\\
\nonumber x\Delta
s(x)&=&-0.02434~x^{0.3837}~(1-x)^{7.649}~(~1+3.656~x^{1/2}+
19.50~x~)~,\\
x\Delta G(x)&=&~~736.1~x^{3.678}~(1-x)^{6.879}~(~1-3.147~x^{1/2}+
3.148~x~)~. \label{Set2NLOJET}
\end{eqnarray}
\vskip 0.6cm {\bf B. Higher twist corrections}

We present here the best fit parametrizations of the extracted
values of higher twist corrections to $g_1$ for the proton and
neutron targets (see Table 2) which are valid in the experimental
$x$ region: $0.005 \leq x \leq 0.75~.$

{\bf Fit} $[g_1(x, Q^2)_{\{\rm LO+TMC\}}+h(x)/Q^2]$:
\setcounter{equation}{0}
\renewcommand\theequation{B.\arabic{equation}}
\begin{eqnarray}
\nonumber h^p(x)&=&0.0223 - {0.2186 \over \sqrt{\pi/2}}
exp[-2((x-0.1905)/0.1761)^2] \\
[2mm] h^n(x)&=&0.0362 + {0.2428 \over \sqrt{\pi/2}}
exp[-2((x-0.0267)/0.1768)^2] \label {HTLOparam}
\end{eqnarray}
\vskip 0.6cm

{\bf Fit} $[g_1(x, Q^2)_{\rm \{NLO{(\rm \overline{MS})}+TMC\}}+
h(x)/Q^2]$:
\begin{eqnarray}
\nonumber h^p(x)&=&0.0360 - {0.1706 \over \sqrt{\pi/2}}
exp[-2((x-0.2030)/0.1736)^2] \\
[2mm] h^n(x)&=&0.0176 + {0.2313 \over \sqrt{\pi/2}}
exp[-2((x-0.0723)/0.1748)^2] \label {HTMSparam}
\end{eqnarray}
\vskip 0.6cm

{\bf Fit} $[g_1(x, Q^2)_{\rm \{NLO(\rm JET)+TMC\}}+ h(x)/Q^2]$:
\begin{eqnarray}
\nonumber h^p(x)&=&0.0373 - {0.1629 \over \sqrt{\pi/2}}
exp[-2((x-0.2046)/0.1700)^2] \\
[2mm] h^n(x)&=&0.0137 + {0.2519 \over \sqrt{\pi/2}}
exp[-2((x-0.0676)/0.1812)^2] \label {HTJETparam}
\end{eqnarray}

\newpage
\begin{center}
\begin{tabular}{cl}
&{\bf TABLE I.} The parameters of the {\bf Set 1} NLO input
polarized PD at $Q^2=1~GeV^2$ \\ & as obtained from the best
'$g1/F1$' fits to the world data in the $\rm \overline{MS}$ and
JET schemes.
\\&The errors shown are total (statistical and systematic).
The parameters marked \\&by (*) are fixed by the sum rules
(\ref{ga}) and (\ref{3FD}).
\end{tabular}
\vskip 0.6 cm
\begin{tabular}{|c|c|c|c|c|c|c|} \hline
    Fit &~~~~$g_1^{\rm NLO}/F_1^{\rm NLO}(\rm \overline{MS})$~~~~
    &~~~~$g_1^{\rm NLO}/F_1^{\rm NLO}(\rm JET)$~~~~\\ \hline
 $\rm DF$        &  190~-~6            &     190~-~6 \\
 $\chi^2$        &  166.2              &     168.3    \\
 $\chi^2/\rm DF$ &  0.903              &     0.915   \\  \hline
 $\eta_u$        &~~0.926$^*$          &    $0.926^*$    \\
 $a_u$           &~~0.207~$\pm$~0.018~~&~~0.206~$\pm$~0.018  \\
 $\eta_d$        &-~0.341$^*$          &    $-0.341^*$      \\
 $a_d$           &~~0.098~$\pm$~0.059~~&~~0.111~$\pm$~0.059  \\
 $\eta_s$        &-~0.061~$\pm$~0.007~~&-~0.051~$\pm$~0.008  \\
 $a_s$           &~~0.637~$\pm$~0.082~~&~~0.595~$\pm$~0.101  \\
 $\eta_g$        &~~0.307~$\pm$~0.175~~&~~0.171~$\pm$~0.166  \\
 $a_g$           &~~2.372~$\pm$~0.691~~&~~3.169~$\pm$~1.323   \\ \hline
\end{tabular}
\end{center}
\vskip 0.6cm

\newpage
\begin{center}
\begin{tabular}{cl}
&{\bf TABLE II.} The parameters of the {\bf Set 2} of LO, NLO($\rm
\overline{MS}$) and NLO(JET) input \\ & polarized PD at
$Q^2=1~GeV^2$ as obtained from the best $(g_1+\rm HT)$ fits to
the \\&world data. The errors shown are total (statistical and
systematic). The \\&parameters marked by (*) are fixed. Note that
the TMC are included in $(g_1)^{\rm LT}$.
\end{tabular}
\vskip 0.6 cm
\begin{tabular}{|c|c|c|c|c|c|c|} \hline
    Fit &~$(g_1)_{\rm LO}^{\rm LT}+h(x)/Q^2$~&~~$(g_1)_{\rm NLO(\rm \overline{MS})}
    ^{\rm LT}+h(x)/Q^2$~~
    &~~$(g_1)_{\rm NLO(\rm JET)}^{\rm LT}+h(x)/Q^2$ \\ \hline
 $\rm DF$        &  190~-~16          &     190~-~16   &  190~-~16  \\
 $\chi^2$        &  156.1             &      154.5     &    154.4 \\
 $\chi^2/\rm DF$ &  0.897             &      0.888     &    0.887 \\  \hline
 $\eta_u$        &~~0.926$^*$         &    $0.926^*$   &     $0.926^*$  \\
 $a_u$           &~~0.000~$\pm$~0.010~&~~0.252~$\pm$~0.037~&~~0.247~$\pm$~0.038 \\
 $\gamma_u$      &~~1.610~$\pm$~0.301~&    $0^*$           &   $0^*$ \\
 $\eta_d$        &-~0.341$^*$         &    $-0.341^*$      &   $-0.341^*$  \\
 $a_d$           &~~0.000~$\pm$~0.062~&~~0.166~$\pm$~0.124~&~~0.143~$\pm$~0.149  \\
 $\gamma_d$      &~~3.797~$\pm$~1.736~&    $0^*$           &   $0^*$  \\
 $\eta_s$        &-~0.068~$\pm$~0.006~&-~0.070~$\pm$~0.008~&-~0.060~$\pm$~0.012 \\
 $a_s$           &~~0.544~$\pm$~0.058~&~~0.656~$\pm$~0.069~&~~0.666~$\pm$~0.114 \\
 $\eta_g$        &~~$0.179^*$~        &~~0.179~$\pm$~0.267~&~~0.231~$\pm$~0.283 \\
 $a_g$           &~~$2.218^*$~        &~~2.218~$\pm$~1.650 &~~2.709~$\pm$~1.567 \\ \hline
 $x_i$           & \multicolumn{3}{|c|}{$h^p(x_i)~[GeV^2]$}   \\  \hline
  0.028          &~-0.010~$\pm$~0.042    &~~0.018~$\pm$~0.047 &~~0.021~$\pm$~0.043  \\
  0.100          &-~0.078~$\pm$~0.034    &-~0.031~$\pm$~0.032 &-~0.023~$\pm$~0.034 \\
  0.200          &-~0.151~$\pm$~0.038    &-~0.100~$\pm$~0.040 &-~0.093~$\pm$~0.043 \\
  0.350          &-~0.011~$\pm$~0.044    &~~0.004~$\pm$~0.046 &~~0.007~$\pm$~0.046 \\
  0.600          &~~0.022~$\pm$~0.021    &~~0.036~$\pm$~0.020 &~~0.037~$\pm$~0.021 \\ \hline
 $x_i$           &    \multicolumn{3}{|c|}{$h^n(x_i)~[GeV^2]$}  \\ \hline
  0.028          &~~0.230~$\pm$~0.061    &~~0.182~$\pm$~0.065 &~~0.195~$\pm$~0.063 \\
  0.100          &~~0.174~$\pm$~0.040    &~~0.196~$\pm$~0.038 &~~0.203~$\pm$~0.038 \\
  0.200          &~~0.064~$\pm$~0.054    &~~0.081~$\pm$~0.061 &~~0.078~$\pm$~0.065 \\
  0.325          &~~0.038~$\pm$~0.026    &~~0.025~$\pm$~0.029 &~~0.023~$\pm$~0.031 \\
  0.500          &~~0.036~$\pm$~0.014    &~~0.014~$\pm$~0.013 &~~0.013~$\pm$~0.015 \\ \hline
\end{tabular}
\end{center}
\vskip 0.6 cm

\end{document}